\documentclass{webofc}
\usepackage[varg]{txfonts}   
\usepackage{graphicx,color,xcolor}
\usepackage{wasysym}
\usepackage[caption=false]{subfig}
\usepackage[colorlinks=True, citecolor=blue, urlcolor=blue, linkcolor=blue]{hyperref}

\begin{document}
\title{Testing Eigenstate Thermalization Hypothesis for Non-Abelian Gauge Theories}
%
%

\author{
\firstname{Xiaojun} \lastname{Yao}\inst{1}\fnsep\thanks{\email{xjyao@uw.edu}} \and
\firstname{Lukas} \lastname{Ebner}\inst{2}\fnsep\thanks{\email{Lukas.Ebner@stud.uni-regensburg.de}} \and
\firstname{Berndt} \lastname{M\"uller}\inst{3}\fnsep\thanks{\email{berndt.mueller@duke.edu}} \and
\firstname{Andreas} \lastname{Sch\"afer}\inst{2}\fnsep\thanks{\email{andreas.schaefer@physik.uni-regensburg.de}} \and
\firstname{Clemens} \lastname{Seidl}\inst{2}\fnsep\thanks{\email{Clemens.Seidl@physik.uni-regensburg.de}} 
}

\institute{
InQubator for Quantum Simulation, Department of Physics,
University of Washington, Seattle, Washington 98195, USA
\and
Institut f\"ur Theoretische Physik, Universit\"at Regensburg, D-93040 Regensburg, Germany
\and
Department of Physics, Duke University, Durham, North Carolina 27708, USA
}

\fnsep{\!\!IQuS@UW-21-068}

\abstract{We report on progress in full quantum understanding of thermalization in non-Abelian gauge theories. Specifically, we test the eigenstate thermalization hypothesis for (2+1)-dimensional SU(2) lattice gauge theory. 
}

\maketitle
\section{Introduction}
\label{intro}
How highly excited states in QCD thermalize is an interesting theoretical question, relevant for heavy ion collisions. Previous studies have mainly used classical or semiclassical methods, which omit or approximate essential properties of quantum mechanical systems. An alternative paradigm of understanding thermalization of an isolated quantum system is the eigenstate thermalization hypothesis (ETH), which states that matrix elements of local observables $\cal{O}$ in the energy eigenstate basis can be written as (see a recent review~\cite{DAlessio:2015qtq})
\begin{align}
\langle E_\alpha |{\cal O}| E_\beta \rangle = \langle {\cal O} \rangle_{\rm mc} (E)\delta_{\alpha\beta} + e^{-S(E)/2} f_{\cal O}(E,\omega) R_{\alpha\beta} \,,\!
\label{eq:ETH}
\end{align}
where the diagonal part is the microcanonical ensemble expectation value and the off-diagonal part is exponentially suppressed by the system's entropy $S(E)$ with a modulation given by the spectral function $f_{\cal O}(E,\omega)$. Here $E = (E_\alpha + E_\beta)/2$, $\omega = E_\alpha - E_\beta$ and the $R_{\alpha\beta}$ values vary radically with zero mean and unit variance. Under the ETH, one can show that the system will look like thermal when probed through the observable ${\cal O}$ after some time, even though the whole system is still a pure state.

It is generally expected that nonintegrable systems obey ETH, and so do non-Abelian gauge theories. However, an explicit demonstration of the ETH for non-Abelian gauge theories is still missing. Here we report on progress in testing the ETH for (2+1)-dimensional SU(2) lattice gauge theory by numerical exact diagonalization of three lattice setups: (a) a linear plaquette chain with the electric field basis limited to $j=0,\frac{1}{2}$, (b) a two-dimensional hexagonal lattice with the same Hilbert space constraint, and (c) a chain of only three plaquettes but such a sufficiently large electric field Hilbert space ($j \leq \frac{7}{2})$ that convergence of all energy eigenvalues in the analyzed energy window is observed.


\section{Hamiltonian of 2+1$D$ SU(2) lattice gauge theory}
\label{sec-H}
\begin{figure*}
\centering
\includegraphics[width=4.6cm,clip]{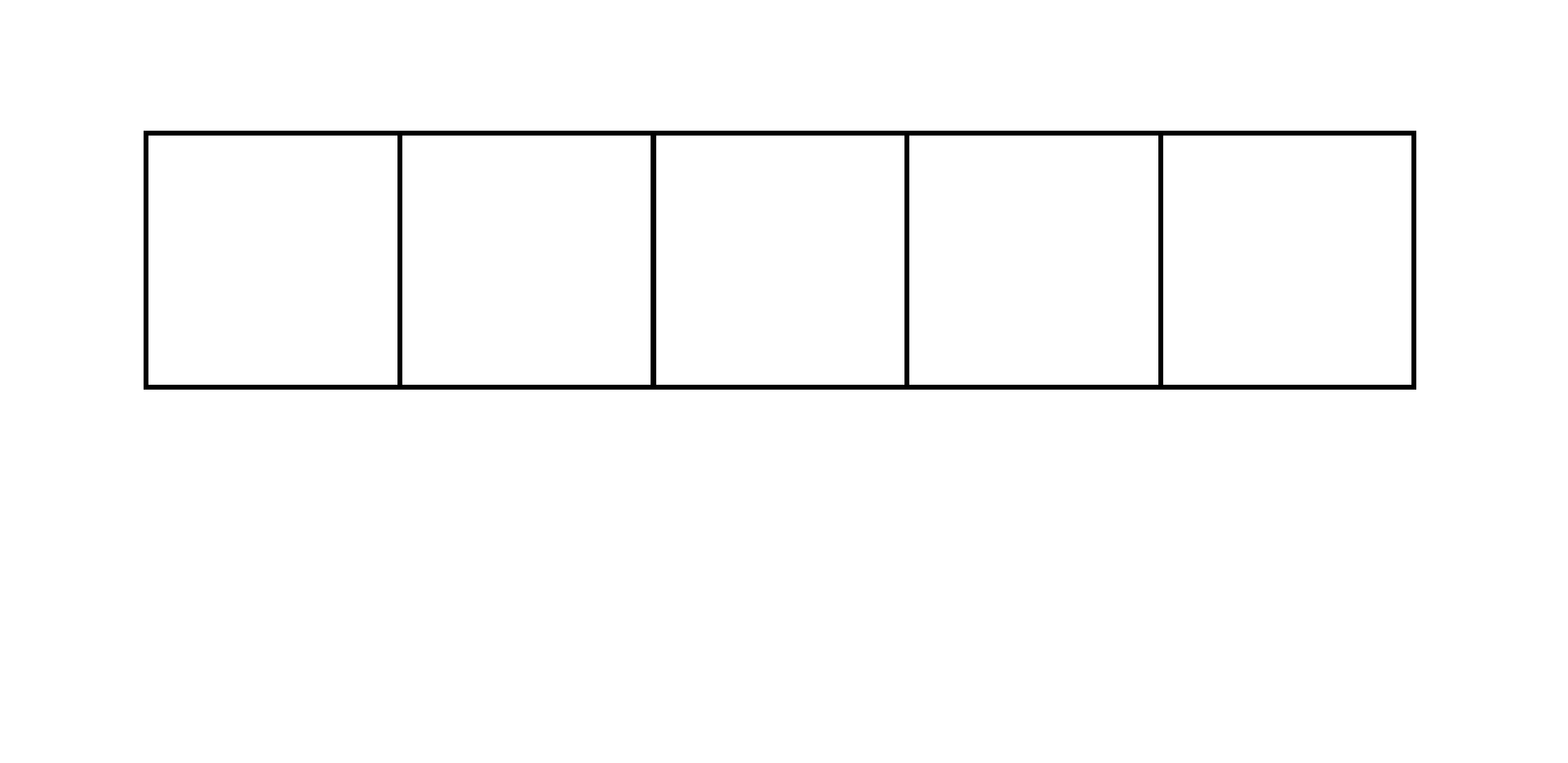}
~~~~~~~
\includegraphics[width=4cm,clip]{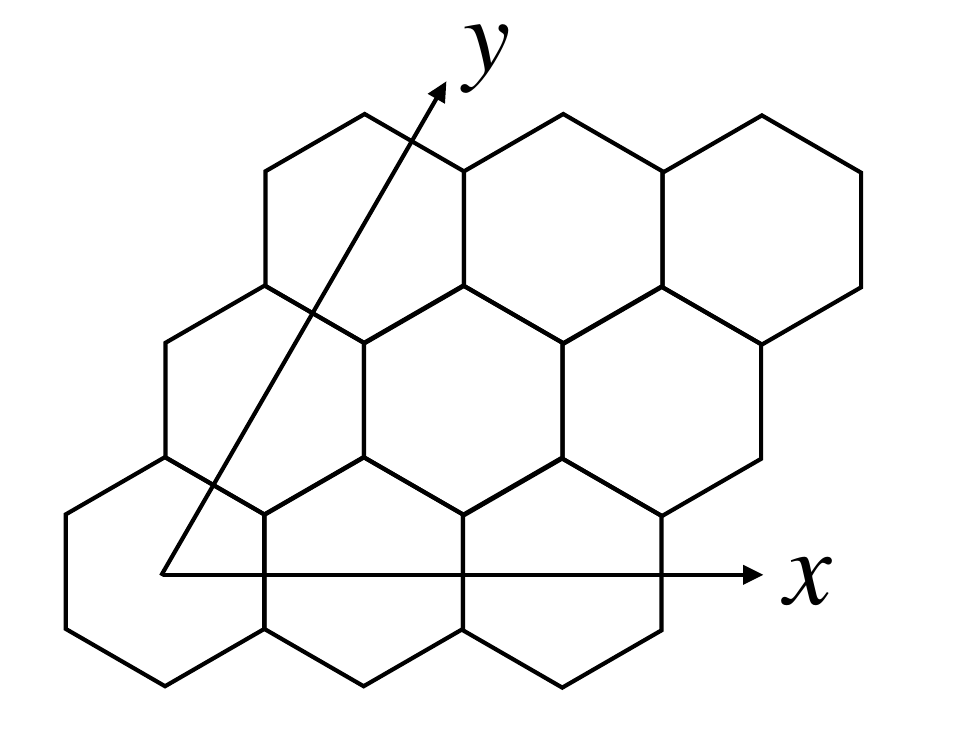}
\vspace*{-0.3cm}
\caption{Examples of lattice shapes considered: a linear chain of 5 square plaquettes (left) and a $3\times3$ plane of hexagonal plaquettes (right).}
\label{fig-lattice}
\end{figure*}

In our studies, we consider two different lattice shapes: a linear chain of square plaquettes and a hexagonal plaquette plane, as shown in Fig.~\ref{fig-lattice}. Throughout this article, we assume periodic boundary conditions. The discretized Kogut-Susskind Hamiltonian can be written as
\begin{align}
\label{eq:H}
    \!H=\frac{g^2}{2}\sum_{\rm links}\sum_{a=1}^3 (E^a)^2-\frac{2}{a^2g^2}\sum_{\rm plaquettes}\Box  \,,\quad H =   \frac{3\sqrt{3}g^2}{4}  \sum_{\rm links} \sum_{a=1}^3  (E^a)^2
- \frac{8\sqrt{3}}{9 g^2a^2}  \sum_{\rm plaquettes} \varhexagon \,,
\end{align} 
for square lattices ($a$ is the square side length) and hexagonal lattices ($a$ is the honeycomb side length) respectively. $\Box$ is the trace of the product of four Wilson lines (link variables) along the four edges of a square plaquette while $\varhexagon$ is that of six Wilson lines along the edges of a honeycomb plaquette.

We use the electric basis to describe states on links. If a link is in $|j\rangle$ state, its electric energy is give by
$\sum_{a=1}^3(E^a)^2 |j\rangle = j(j+1) |j\rangle$.
Physical states satisfy Gauss's law at each vertex, which means they transform as SU(2) singlets. The advantage of a hexagonal lattice in 2D is that physical states at a vertex are uniquely determined by the three $j$ values on the three links attached to the vertex. In the electric basis, matrix elements of the plaquette terms between physical states can be worked out for both the linear chain and the hexagonal lattice.

With a given lattice size and truncation of the local electric basis (i.e., $j_{\rm max}$), the physical Hilbert space is finite and the matrix representation of the Hamiltonian can be obtained, which allows one to exactly diagonalize the system and study its various properties. The calculations can be greatly simplified when $j_{\rm max}=\frac{1}{2}$ since a map into spin systems can be constructed~\cite{Yao:2023pht,Muller:2023nnk}. For calculations of physical observables, one must take the continuum and infinite volume limits, as well as remove the electric basis truncation.

\section{Level statistics in energy spectrum}
\label{sec-level}
The energy spectrum of a quantum system whose classical counterpart is chaotic exhibits level repulsion: The energy gap between nearest states obeys the Wigner-Dyson statistics featuring a vanishing distribution at zero energy gap, rather than the Poisson statistics, in which the distribution at zero energy gap is nonvanishing.

\begin{figure*}
\centering
\subfloat[\label{fig:a}]{%
\includegraphics[width=4cm,clip]{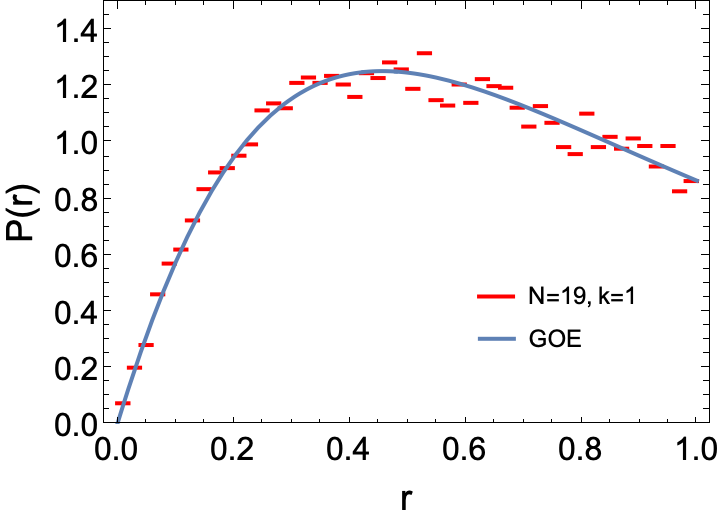}%
}\hfill
\subfloat[\label{fig:b}]{%
\includegraphics[width=4cm,clip]{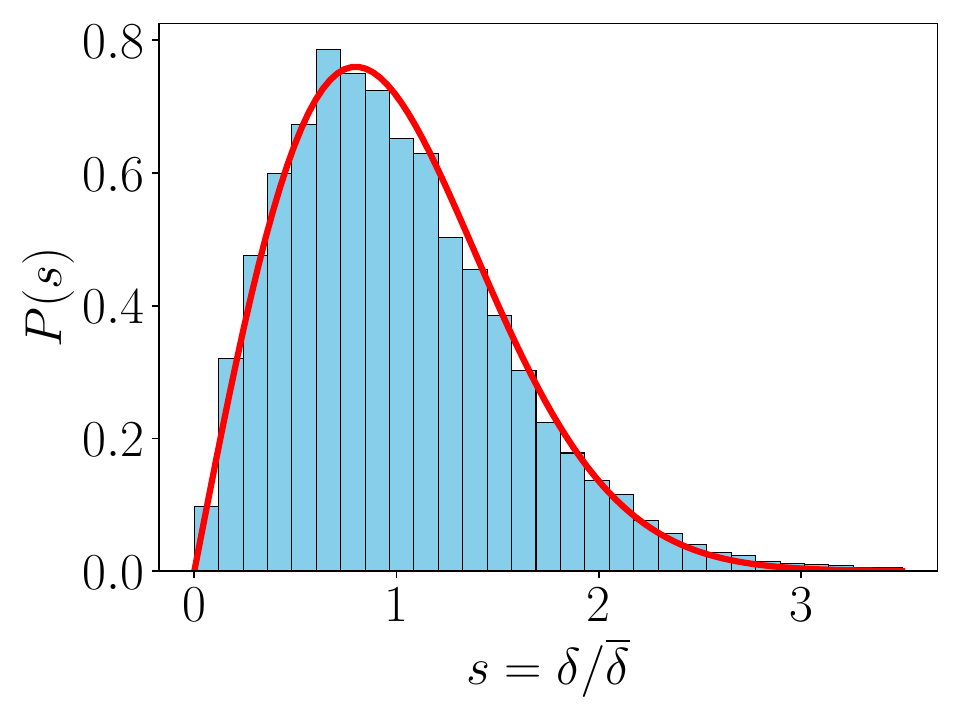}%
}\hfill
\subfloat[\label{fig:c}]{%
\includegraphics[width=4cm,clip]{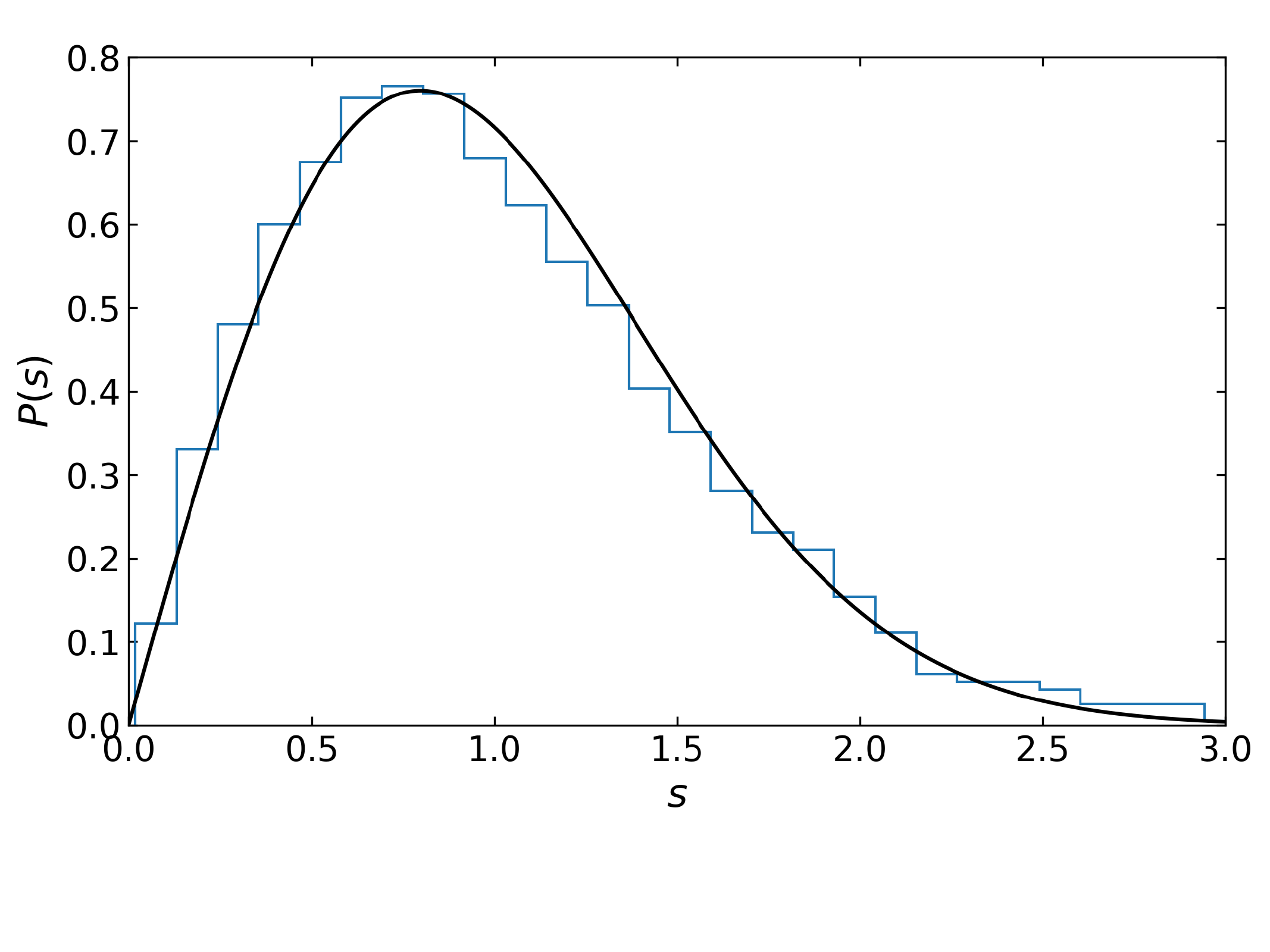}%
}
\vspace*{-0.2cm}
\caption{Energy level statistics of the three lattice setups considered: (a) restricted gap ratio distribution in the $k=\frac{2\pi}{N}$ momentum sector on a $N=19$ plaquette chain with $j_{\rm max}=\frac{1}{2}$, (b) energy gap distribution in the $k_x=\frac{2\pi}{N_x},k_y=\frac{2\pi}{N_y}$ momentum sector on a $N_x\times N_y=5\times4$ hexagonal lattice, where the directions of $x$ and $y$ are as shown in Fig.~\ref{fig-lattice}, and (c) gap distribution of converged energies in the zero momentum sector with positive left-right and top-bottom reflection parity on a $N=3$ plaquette chain with $j_{\rm max}=\frac{7}{2}$~\cite{Ebner:2023ixq}. The distributions are compared with GOE predictions depicted in solid lines.}
\label{fig-gap}
\end{figure*}

From exact diagonalization, we obtain the eigenenergies $E_\alpha$ and compute their nearest gaps $\delta_\alpha = E_{\alpha+1}-E_\alpha$. We also calculate the restricted gap ratios $ r_\alpha \equiv \frac{{\rm min}[\delta_\alpha,\delta_{\alpha-1}]}{{\rm max}[\delta_\alpha,\delta_{\alpha-1}]} $
as an additional sensitive measure of the level statistics. Some of our results are shown in Fig.~\ref{fig-gap}, where we also compare with predictions from the random matrix theory Gaussian orthogonal ensemble (GOE). Our results are consistent with the GOE predictions within statistical uncertainties, which implies our systems are chaotic and expected to obey the ETH.

The analyses are performed with a fixed coupling: $g^2=1.2$ for setup (a), $g^2=0.75$ for setup (b) and $g^2=0.8$ for setup (c) in lattice units. At a fixed lattice size which is limited by our computing resources, certain choices of coupling values work better to demonstrate the ETH. However, we expect any nonzero value will suffice when the lattice is big enough. We will use the same set of coupling values in the following.

\section{Diagonal matrix elements}
\label{sec-diag}
\begin{figure*}
\centering
\includegraphics[width=5cm,clip]{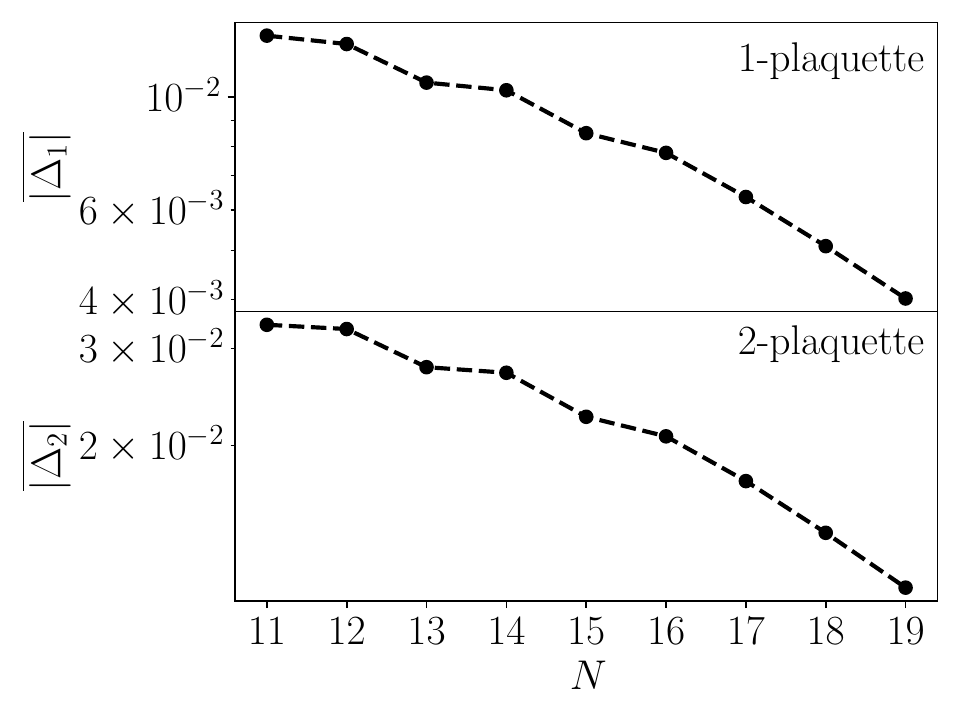}
~~~~~~~~~
\includegraphics[width=5cm,clip]{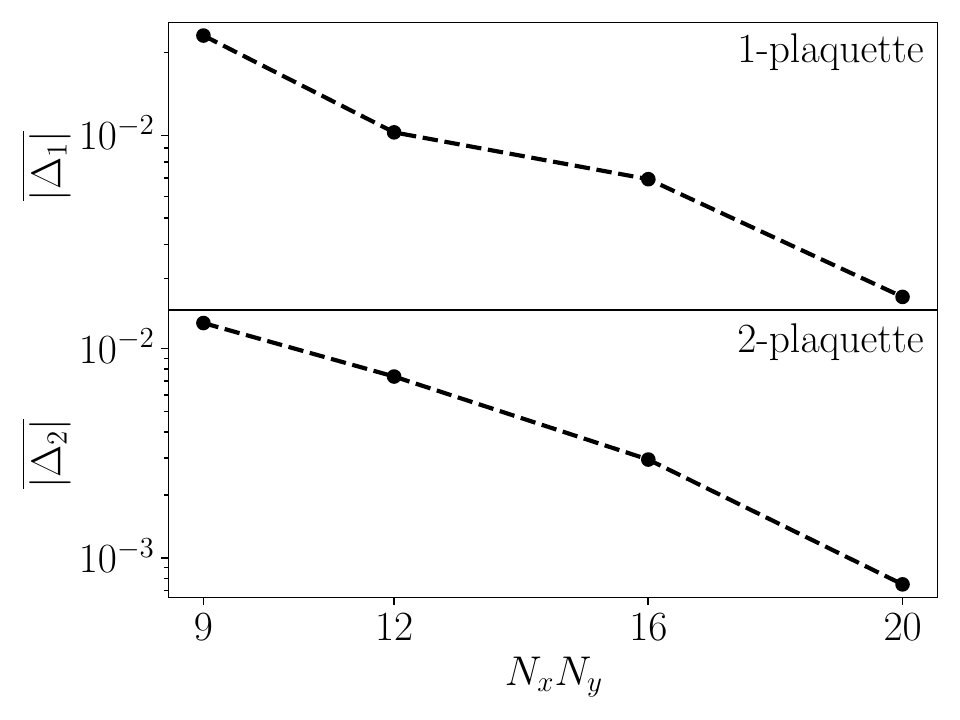}
\vspace*{-0.3cm}
\caption{Averaged deviation of the diagonal matrix element from the microcanonical ensemble for varying lattice sizes. Left: linear chain with $j_{\rm max}=\frac{1}{2}$~\cite{Yao:2023pht}. Right: honeycomb lattice with $j_{\rm max}=\frac{1}{2}$~\cite{Ebner:2023ixq}.}
\label{fig-diag}
\end{figure*}

We then study how much the diagonal matrix elements of local operators such as 1-plaquette ${\cal O}_1$ and 2-plaquette ${\cal O}_2$ operators differ from the microcanonical ensemble expectation values. To this end, we introduce a proxy of the microcanonical ensemble expectation value by averaging over matrix elements of nearby 21 states and define the deviation as $
\Delta_i(\alpha) \equiv \langle \alpha| {\cal O}_i |\alpha\rangle - \frac{1}{21}\sum_{\beta=\alpha-10}^{\alpha+10} \langle \beta| {\cal O}_i |\beta\rangle $.
If the system obeys the ETH, the deviation will decrease exponentially with the system size, since the entropy in Eq.~\eqref{eq:ETH} is extensive. We take the magnitude of $\Delta_i(\alpha)$ and average over all states in the middle of the spectrum and obtain results as shown in Fig.~\ref{fig-diag} for varying lattice sizes. Exponential decreases in the deviation are observed for both the plaquette chain with $j_{\rm max}=\frac{1}{2}$ and the honeycomb lattice with $j_{\rm max}=\frac{1}{2}$. 

\section{Off-diagonal matrix elements}
\label{sec-off}
Finally, we study off-diagonal matrix elements of the electric energy operator $H_{\rm el}$, which is the first term in the Hamiltonian~\eqref{eq:H}. We first study the magnitudes of the off-diagonal matrix elements $|\langle \alpha|H_{\rm el}| \beta\rangle |$ and find they follow Gaussian distributions in small $\omega$ windows with a given $E$. By either calculating the second moment of the distribution or fitting with a Gaussian function, we obtain the width $\sigma$ of the Gaussian distribution, which allows us to calculate the spectral function via
$f_{\rm el}(E,\omega) = \sigma \sqrt{\rho(E)}$, where $\rho(E) =e^{S(E)}$ is the state density at energy E.

\begin{figure*}
\centering
\includegraphics[width=4.7cm,clip]{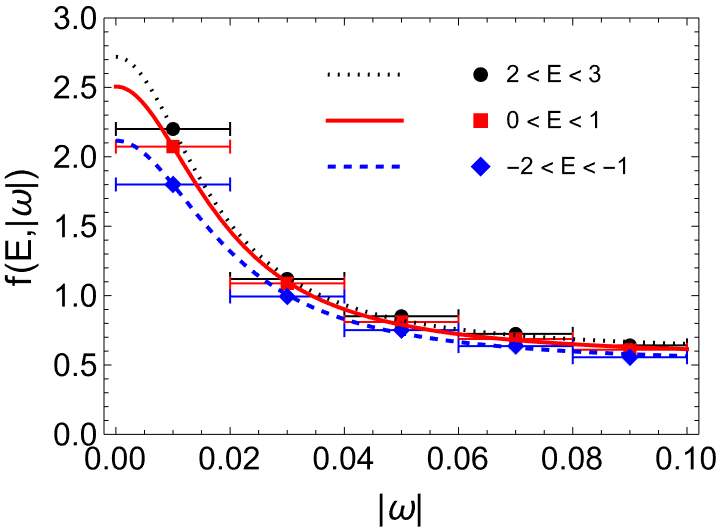}
~~~~~~~~~
\includegraphics[width=5cm,clip]{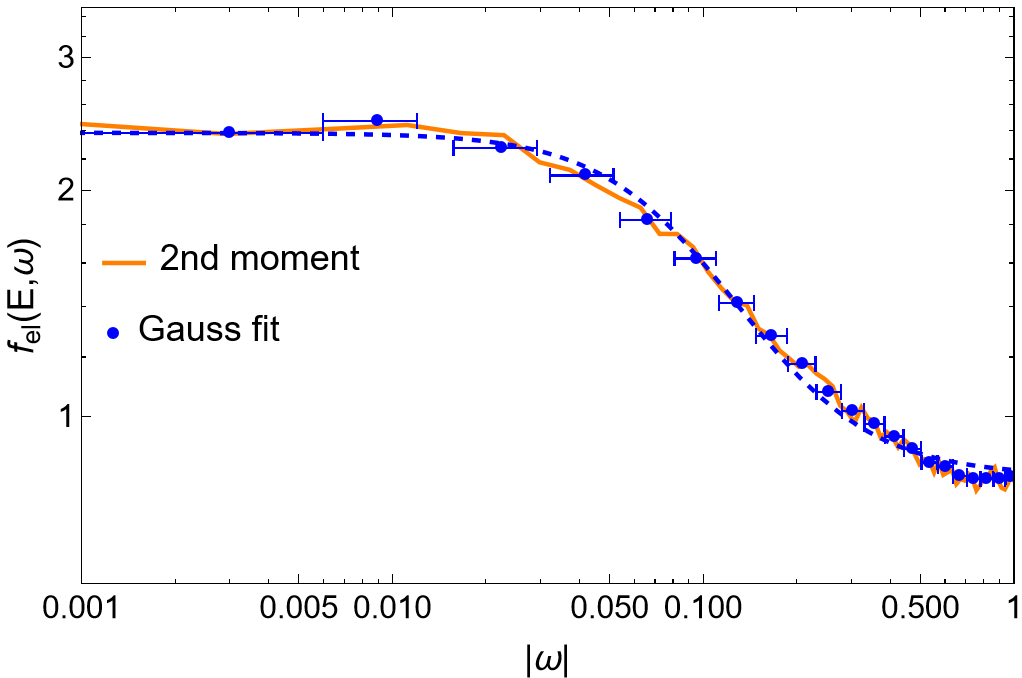}
\vspace*{-0.3cm}
\caption{Spectral function of the electric energy operator in several energy windows for the plaquette chain with $N=19$, $j_{\rm max}=\frac{1}{2}$ (left) and $N=3$, $j_{\rm max}=\frac{7}{2}$ (right). The lines on the left and the dashed line on the right are fits with Lorentzian functions~\cite{Ebner:2023ixq}.}
\label{fig-off}
\end{figure*}

Our results are shown in Fig.~\ref{fig-off} for setup (a) and (c), where we see the spectral function in each case develops a transport peak at small $\omega$, which can be well-described by a Lorentzian shape. 

Future studies will be extended to the SU(3) case, higher dimensions and the inclusion of dynamical fermions. We will also study the onset of GOE behavior shown in the off-diagonal matrix elements in more detail~\cite{Ebner:2023ixq}.

X.Y. was supported in part by the U.S. Department of Energy, Office of Science, Office of Nuclear Physics, InQubator for Quantum Simulation (IQuS) (https://iqus.uw.edu) under Award Number DOE (NP) Award DE-SC0020970 via the program on Quantum Horizons: QIS Research and Innovation for Nuclear Science. B.M. acknowledges support from the U.S. Department of Energy Office of Science (Grant DE-FG02-05ER41367) and from Yale University during extended visits. The authors gratefully acknowledge the scientific support and HPC resources provided by the Erlangen National High Performance Computing Center (NHR@FAU) of the Friedrich-Alexander-Universität Erlangen-Nürnberg (FAU) under the NHR project ID b172da. NHR funding is provided by federal and Bavarian state authorities. NHR@FAU hardware is partially funded by the German Research Foundation (DFG) – 440719683. This work was partially facilitated through the use of advanced computational, storage, and networking infrastructure provided by the Hyak supercomputer system at the University of Washington.

\end{document}